# Switching Casimir forces with Phase Change Materials

G. Torricelli[1], P. J. van Zwol[2], O. Shpak[2], C. Binns[1], G. Palasantzas[2], B.J. Kooi[2]

V. B. Svetovoy[3], M. Wuttig[4]

[1] Department of Physics and Astronomy, University of Leicester, Leicester LE1 7RH, United Kingdom

[2] Materials innovation institute M2i and Zernike Institute for Advanced Materials, University of Groningen, 9747 AG Groningen, The Netherlands

[3] MESA+ Institute for Nanotechnology, University of Twente, PO 217, 7500 AE Enschede, The Netherlands

[4] I. Physikalisches Institut (IA), RWTH Aachen University, 52056 Aachen, Germany

**Abstract**

We demonstrate here a controllable variation in the Casimir force. Changes in the force of up to 20% at separations of ~100 nm between Au and AgInSbTe (AIST) surfaces were achieved upon crystallization of an amorphous sample of AIST. This material is well known for its structural transformation, which produces a significant change in the optical properties and is exploited in optical data storage systems. The finding paves the way to the control of forces in nanosystems, such as micro- or nanoswitches by stimulating the phase change transition via localized heat sources.

Pacs numbers: 78.68.+m, 03.70.+k, 85.85.+j, 12.20.Fv

Casimir forces [1-8] arise between two surfaces due to the quantum zero-point energy of the electromagnetic field. The surfaces restrict the allowed wavelengths and thus the number of field modes within the cavity, which locally depresses the zero point energy of the electromagnetic field. The reduction depends on the separation between the plates thus there is a force between them, which for normal materials is always attractive [1]. The zero point energy manifests itself as quantum fluctuations, which in the small separation limit give rise to the familiar van der Waals force. The original calculation of the Casimir force assumed two parallel plates with an infinite conductivity [1]. This was later modified to include the dielectric properties of real materials and the intervening medium [2, 3], providing the first glimpse of possible methods to control the magnitude and even the direction of the force. This finding has motivated our attempts to manipulate the dielectric properties of a material and hence generate force contrast [9-11]. A particularly exciting possibility is to produce a 'switchable' force by employing materials whose optical properties can be changed *in situ* in response to a simple stimulus [9, 10]. So far the only significant contrast that has been demonstrated is only between different materials [11]. To obtain a large Casimir force contrast for a single material requires a large modification of its dielectric response, which has not been achieved in materials used up to now.

Here we demonstrate that phase change materials (PCMs) [12-21], which are renowned to switch reporducibly between an amorphous and a crystalline phase, are very promising candidates to achieve a significant force contrast without a change of composition. These materials are already used in rewriteable optical data storage [13, 14, 23-25], where the pronounced optical contrast between the amorphous and crystalline state is employed to store information. This storage principle employs a focussed laser beam to locally heat a disk with a thin film of phase change material. Upon a variation of the power and length of the laser pulse the material can be reversibly switched between the amorphous and the crystalline phase many times. Here we will show that the pronounced contrast of optical properties enables a significant change of the Casimir force upon the phase transformation, not previously found in any material [9, 10]. The good cyclability of phase change materials ensures the realization of a switchable Casimir force device.

In order to measure Casimir forces in PCMs, we prepared 1 µm thick amorphous AgInSbTe (AIST) thin films onto standard Al coated Si wafers, of which half of the AIST films were annealed to the crystalline state. The samples were optically characterized by ellipsometry in the frequency range $\omega$=0.04-8.9 eV (see Fig. 1). For the crystalline sample the ellipsometry measurements were directly inverted to obtain the dielectric function [22]. For the amorphous film, because it is transparent in the infrared (IR) range, the system was modelled as an amorphous film above an optically thick Al substrate. The substrate optical properties are important only in IR range, where absorption of the film is very weak. Therefore, it is justified to use tabulated data for the Al substrate.

Since the crystalline film exhibits metallic conductivity, a Drude model was fitted to measured IR data enabling extrapolation below $\omega$<0.04 eV, where data is not available. For the amorphous state this range has an insignificant effect on the force. At high frequencies $\omega$>8.9 eV, where absorption is already small, the imaginary part of the dielectric function $\varepsilon(\omega) = \varepsilon'(\omega) + j\varepsilon''(\omega)$ was extrapolated as $\sim 1/\omega^3$. The extrapolations are justified by a good Kramers-Kronig (KK) consistency for amorphous and crystalline films, and good agreement with the permittivities of the films found previously [23]. As can be seen from Fig. 1, the transformation from the amorphous to the crystalline state leads to drastic changes of the optical properties. These pronounced changes have been recently attributed to a change of bonding upon crystallization [13, 14, 23]. The large change of the dielectric function upon crystallization suggests that a significant change in the Casimir force should be observed.

The measured dielectric response allows Casimir force calculations using the Lifshitz theory (Fig. 2) [2, 3], for which the force depends on the dielectric function at imaginary frequencies (inset Fig. 1). However, such forces are also affected by the surface roughness. The typical roughness of the samples was a few nm rms, but with a few isolated local peaks as evidenced by atomic foce microscopy (AFM) analysis (lower inset in Fig.2). This small roughness is negligible for the Casimir force calculation at separations above 70 nm [26].

The Casimir force measurements, as in Fig. 2, were performed using dynamic AFM mode within an ultra high vacuum (UHV) Atomic Force Microscope (Omicron VT STM/AFM) [27, 28]. Forces were measured in the sphere-plate geometry between a Au coated (100 nm thick) sphere 20.2 μm in diameter, attached to the end of a cantilever. The latter initially vibrates at its resonant frequency, 83.6 kHz, far from the surface. As the sphere approaches the PCM surface, we measure the frequency shift induced by the sphere-plate interaction, which is proportional to the force gradient in the linear approximation. Each experimental force curve is an average of 13 measurements taken in different areas on both samples.

The force measurement method and the experimental set-up are described in detail in [28]. Indeed, precise comparison of force measurements with theory is only possible if we determine electrostatically several, a priori unknown, parameters such as the starting separation distance $Z_0$ for the force measurement (corresponding here to the shortest separation), the cantilever spring constant k, and the contact potential difference $V_0$ [28]. The calibration is performed by measuring the force gradient versus separation distance for two different applied bias voltages $V_b$ on the sphere yielding a gap voltage $\Delta V=V_b-V_0$. The contact potential $V_0$ may not be constant [11, 27, 29] but instead can depend on the separation distance Z between sphere and sample surface. Prior to force acquisition, the determination of $V_0$ is performed at only one distance $Z_0=42.8\pm0.5$ nm for the amorphous, and $Z_0=42.9\pm0.4$ nm for the crystalline phase sample. Then we define $V_0=0$ at $Z=Z_0$ as the reference potential, and the two values are chosen for $V_b$ (-0.5 V, +0.5 V) to obtain the electrostatic force curves. Determination of $Z_0$ and k is achieved by fitting the average of these two force measurements after subtraction of the Casimir contribution (measured for $V_b=0$ V), without the calibration being affected by variations in $V_0$. [28]. The fit gives consistent spring constants, namely, $k=10.8\pm0.3$ N/m for the amorphous film, and $k=10.7\pm0.3$ N/m for the crystalline film.

The experimental uncertainty in the force measurement as deduced from the standard deviation of the cantilever spring constant k and the starting separation distance $Z_0$ is about 7% for both samples. Therefore, the upper inset in Fig.2 demonstrates unambiguously that the gradient of the Casimir force increases in magnitude by approximately 20 % as a result of the transition from the amorphous to the crystalline state. Both the size and the sign of this force change upon crystallization are in qualitative agreement with the theoretical calculations. At short separations (< 55 nm) the increase in the difference is most likely to be attributed to the larger roughness of the crystalline state (lower inset Fig.2). leading to a larger force [26].

The theory based on the measured optical properties predicts a force smaller than the measured one by 8-18 %. The deviation is smaller for the amorphous sample but in both cases it is larger than the experimental and theoretical errors. This deviation cannot be explained by a vertical drift of the AFM probe since the feedback loop maintains the sphere at separation $Z_0$ from the surface (positioning accuracy better than ~0.1 nm). In addition it cannot be explained by the fact that the electrostatics have been performed using an approximate formula for capacitance gradient [28] which leads to an error of $Z_0$ of ~0.2 nm. Also, in order to check the force measurements we used a sample coated with low roughness Au (~1 nm rms) and it was found close agreement between the measured and theoretically predicted forces. Possible uncertainties in the optical properties of PCM due to low and high frequency extrapolations, variation of the substrate properties or film thickness are excluded since they have small influence on the force calculation.

Hence the observed deviation between theory and experiment can be attributed to surface roughness as discussed recently in [30]. Indeed, the electrostatic force involves a larger interaction area on the plate than the Casimir force [30]. Larger areas contain more high peaks so that the averaged surface of the plate will be located higher than for smaller areas [30]. This is specific to the PCM roughness as the inset in Fig. 2 shows. As a result the absolute separation as determined from the electrostatic calibration underestimates the separation in the case of the Casimir interaction. This difference can be ~1-2 nm [30], and it is smaller for the amorphous film. In fact, if the experimental force data are shifted to the left by 1-2 nm, the agreement with the theory is restored.

It is observed that there is a residual electrostatic force $\sim V_0(Z)^2$, where $V_0$ is the sphere-plate contact potential difference [28], which must be subtracted from the measured force. This is possible if $V_0(Z)$ is known for all separations Z used for the force measurements. The variation of $V_0(Z)$ can be extracted from the two electrostatic measurements (applied potentials $V_b=\pm 0.5V$) by simple data manipulations (Fig. 3a) [28]. Variations for $V_0$ between 0-20 mV were observed for separations 40-150 nm without significant differences between amorphous and crystalline samples. As the inset in Fig. 3a indicates, subtraction of this residual electrostatic contribution corresponds to a correction of 6 % at Z=150 nm, and much less than 1% at Z=50 nm as compared to the Casimir force. Therefore, even avoiding this correction, the contrast of the force gradient between the two phases (inset Fig. 2) would remain practically unaffected. Finally, in order to fully confirm our force measurements, another electrostatic measurement was performed under identical conditions as before but with $V_b$=-50 mV (Fig. 3b). Again comparison of the force measurements with the theory, using the parameters extracted from the electrostatic calibration, shows very good agreement. Notably, as Fig. 3b shows, the agreement is better than that for the Casimir force measurements even though the force gradient for $V_b$=-50 mV is much smaller confirming that the thermal drift is well compensated by the feedback loop.

In conclusion, as expected from the pronounced difference in the dielectric function of the amorphous and crystalline phases in phase change materials, a significant difference in the measured Casimir force between the PCM and Au is found for the two states. The measured force contrast is the largest reported to date for a switchable material [9, 10]. Although switching a large area of PCM requires high currents, when the nanometer regime is entered modest currents are sufficient to switch the PCM material. Indeed, the smaller the PCM cell the faster it can switch [24]. Switching times of a few nanoseconds already render phase change materials as very useful in electronic and optical memory applications [24]. Currently, there is a continuing effort to improve the number of switching cycles up to $10^{15}$ making for example PRAM (phase change random access memory) suitable to replace DRAM (Dynamic random access memory) [20]). The property portfolio of suitable dielectrical properties, fast switching, good scalability down to the nanometer regime [24], and strong Casimir force contrast deem PCMs to be promising candidates for a switchable Casimir force device.

**Acknowledgments**

The research was carried out under project number MC3.05242 in the framework of the Research programme of the Materials innovation institute M2i. The force measurements were supported by the UK EPSRC grant EP/F035942/1, and the ESF/CASIMIR grant 3108. The careful preparation of the samples by Michael Woda and Stephan Kremers is gratefully acknowledged. Finally, the authors benefited from exchange of ideas by the ESF Research Network CASIMIR.

**Figure captions**

**Figure 1** (Color online) Absorptive part of the dielectric function for the crystalline (triangles) and amorphous (solid line-circles) state of the AIST film obtained with ellipsometry as a function of frequency. The inset shows the same dielectric functions at imaginary frequencies $\zeta$, which are necessary for the Casimir force calculations using Lifshitz theory.

**Figure 2** (Color online) Casimir force gradient measurement for the crystalline ($\Delta$) and amorphous (•) phase. The calculcated force gradient are depicted as solid and dashed lines for the crystalline and amorphous phase, respectively. The upper inset shows the relative difference between the two force states, normalized with respect the amorphous state, for both the experimental (•) and theoretical (—) data. This inset demonstrates clearly that theory can qualitatively reproduce the measured difference of the Casimir force. The lower inset shows an AFM topography of amorphous (left) and crystaline (right) films.

**Figure 3** (Color online) **a)** Determination of $V_0$ for the crystalline ($\Delta$) and amorphous (•) phase. The inset shows the force contribution due to varying $V_0$, divided by the Casimir force in percent. The latter indicates how much the remaining electrostatic interaction would contribute to the Casimir force if it was not subtracted. **b)** Comparison between the experimental electrostatic force gradient for $V_b$=-50mV (•) and the theory taking into account the measured $V_0$ variations i.e. for $\Delta V$=-50 mV-$V_0(z)$ (—) and without i.e. $\Delta V$=-50 mV (---).

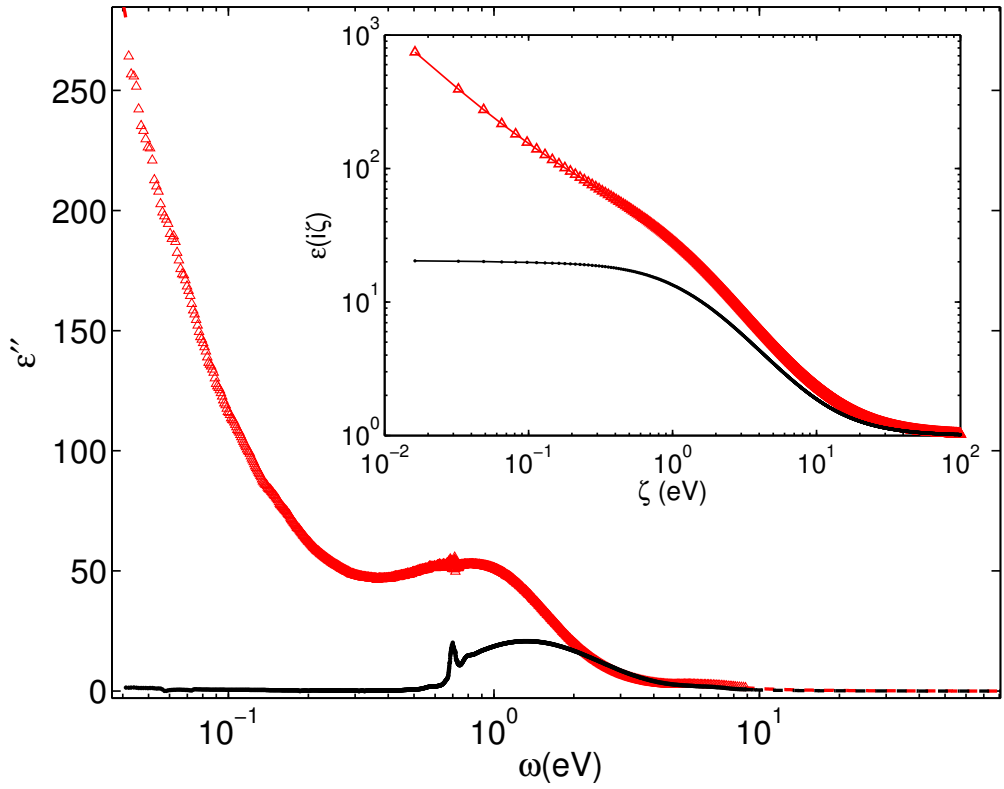

**Figure 1**

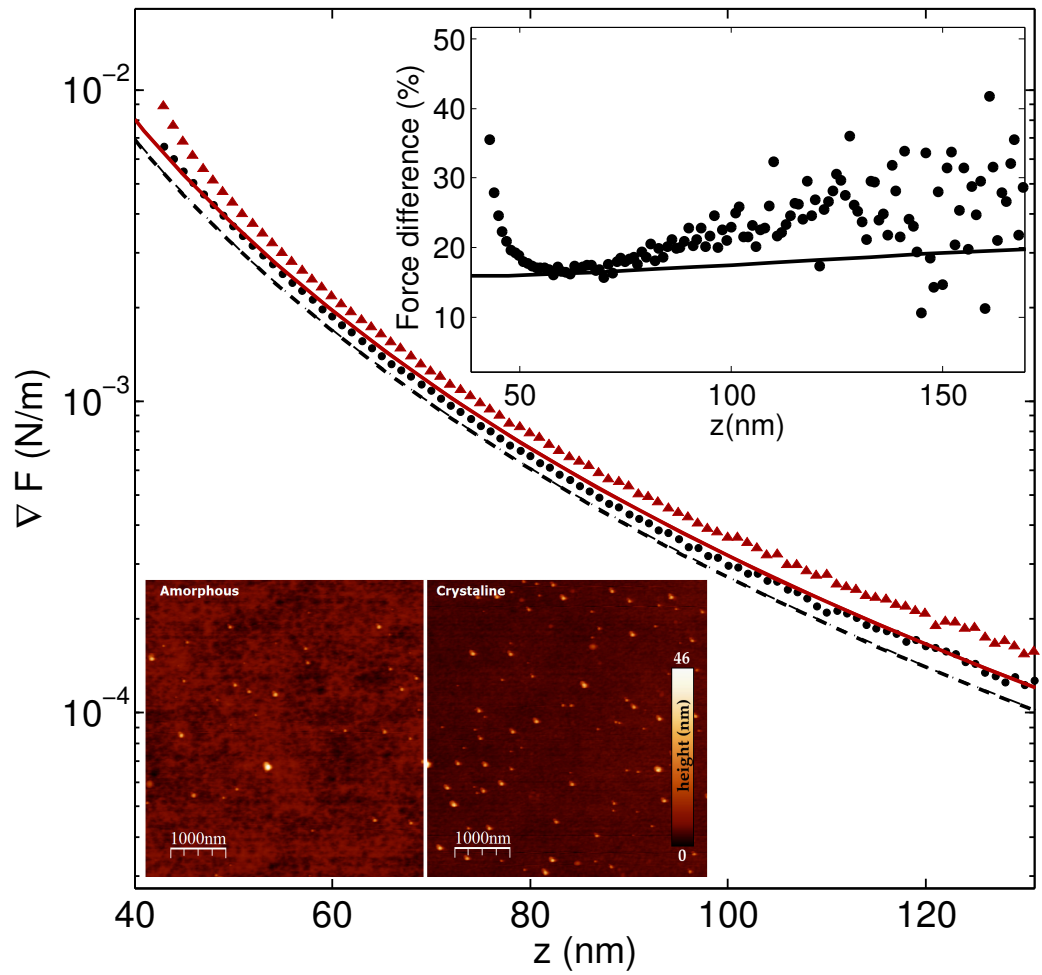

**Figure 2**

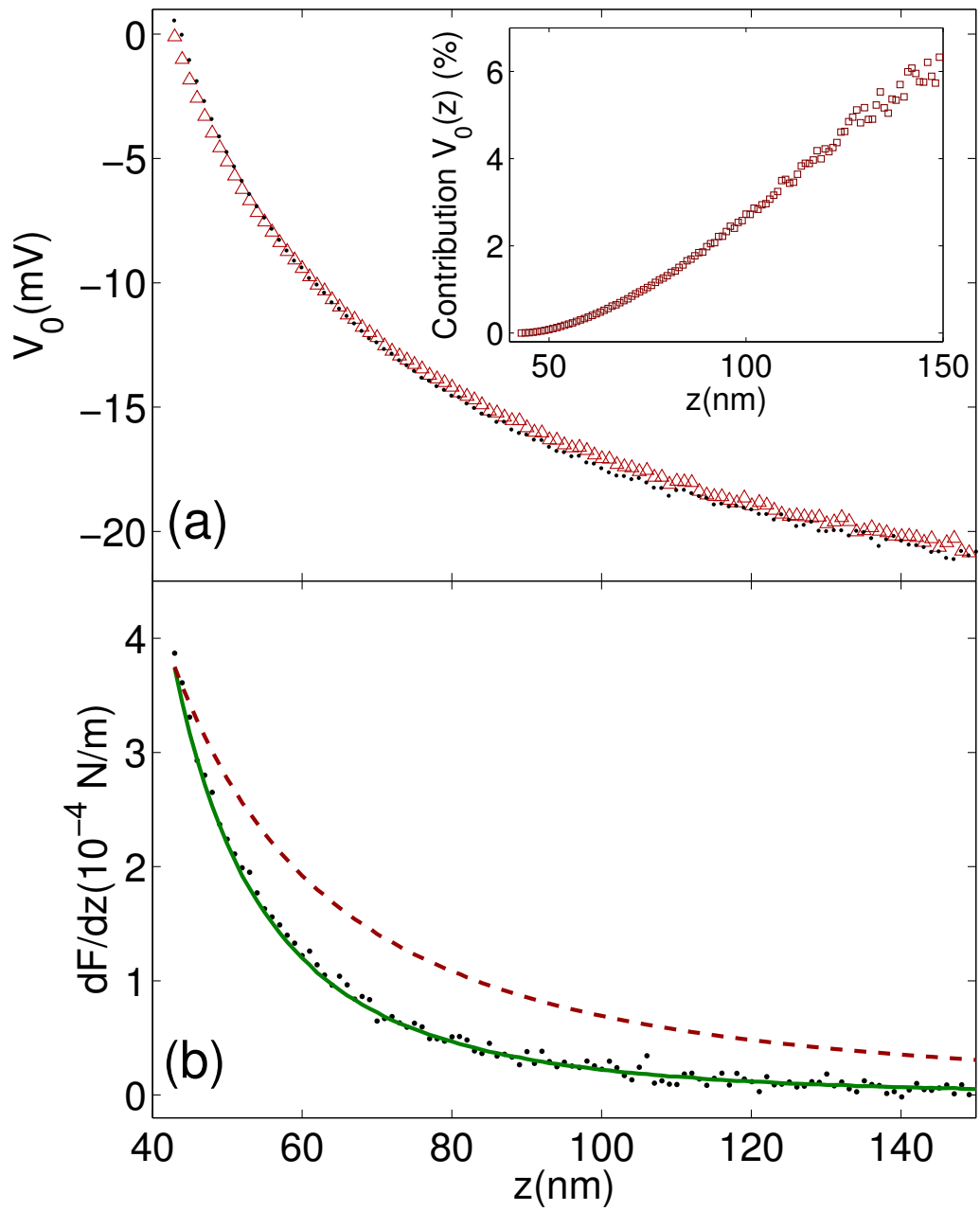

Figure 3